# A method of laser frequency stabilization based on the effect of linear dichroism in alkali metal vapors in a modulated transverse magnetic field

*M.V. Petrenko[1], A.S. Pazgalev[1], and A.K. Vershovskii[1*]*

[1]*Ioffe Institute, Russian Academy of Sciences, St. Petersburg, 194021 Russia*
[*]*e-mail address: antver@mail.ioffe.ru*

We present a method of laser frequency stabilization based on the linear dichroism signal in a transverse magnetic field. The method is similar to the DAVLL (Dichroic Atomic Vapor Laser Lock) method. It differs from DAVLL and from its existing modifications primarily by the fact that it uses signal of linearly polarized light caused by alignment, rather than circular refraction caused by orientation, and therefore allows to obtain error signals at the magnetic field modulation frequency (or its second harmonic) by extremely simple means. The method does not require the strong magnetic fields or careful shielding of the working cell. The method allows the laser frequency to be stabilized in the vicinity of the low-frequency transition in the $D_1$ line of Cs. Although the absorption line in a gas-filled cell is typically gigahertz wide, the achievable resolution, limited by the signal-to-noise ratio of photon shot noise, can reach tens of kilohertz in one hertz bandwidth.

**Keywords:** *laser frequency stabilization, dichroic atomic vapor laser lock, linear dichroism, optical alignment, linearly polarized light, transverse magnetic field.*

## I. INTRODUCTION

The development of new precise spectroscopic methods and closely related methods of quantum optics and spintronics has led to the emergence of new areas of applied physics. New optical methods of controlling spin states of atomic ensembles, cooling of atoms and molecules [1], and interferometry of cold atoms [2] have resulted in creation of new secure communication systems [3,4], optical [5] and radio frequency standards using thermal and cold atoms, as well as new classes of quantum sensors (e.g., optical sensors of magnetic field [6–10], electric field [11], rotation [12], temperature [13] and so on). The vast majority of the above methods and devices are based on the use of a highly coherent laser, stabilized in the vicinity of a narrow spectral absorption line in the atomic medium. As a rule, such a medium consists of hydrogen-like structures characterized by the presence of one electron on the outer shell, namely alkali metal atoms (in most cases – rubidium or cesium) in the gaseous phase or isolated positive ions of alkaline earth metals. The field of applications of quantum optics is continuously expanding – an example is a wide class of sensors based on nitrogen-vacancy centers in diamond [14–16] or devices based on rare earth atoms [17] and their ions [18]. Nevertheless, alkali metals undoubtedly remain the most commonly used medium in all of the above methods and tasks.

The aforementioned advanced methods in quantum optics rely on stabilized single-mode lasers. While various techniques exist for stabilizing laser frequency – including stabilization to optical resonators, to the beat frequency with another laser, and many others – long-term frequency stabilization is invariably achieved by referencing to atomic or molecular transitions. Atomic optical transitions, as natural frequency references, offer unparalleled accuracy and long-term stability, making them indispensable for quantum optics instrumentation. Therefore, many applications, including laser cooling, quantum communication, and atomic interferometry, require lasers stabilized to transitions in separate atomic ensembles. A typical stabilization module consists of a cell with working atoms, a thermostat, a magnetic shield, and an optical system for beam shaping and detection. A well-known example is the COSY unit from Thorlabs, which utilizes saturated absorption spectroscopy resonances.

It should be noted, however, that simpler frequency stabilization methods often require laser frequency modulation, with the modulation amplitude being comparable to the width of the absorption line. To stabilize the laser frequency by atomic transitions, the absorption (or saturated absorption) signal is typically used, which requires light frequency modulation and subsequent lock-in amplification [19]. This method has a significant disadvantage: the laser radiation line is split into a number of spectral components that together form an anti-symmetric linear spectrum. At the same time, the most precise methods in both quantum optics (e.g., cold atom interferometry) and laser spectroscopy (in particular, nonlinear sub-Doppler or Raman spectroscopy) impose stringent requirements on the laser spectrum that are incompatible with such modulation.

For a long time, this contradiction was resolved (and continues to be resolved) by using complex, unreliable, and expensive external acousto-optic modulators. These devices not only shift the laser frequency but also alter its direction, necessitating relatively intricate optical schemes to compensate for this effect.

The situation changed dramatically with the invention of the Dichroic Atomic Vapor Laser Lock (DAVLL) technique, which for the first time enabled laser stabilization to an atomic transition without frequency modulation. Within a few years, DAVLL schemes became ubiquitous in laboratories, and several modifications of this method emerged, overcoming its inherent limitations to varying degrees.

This work presents a stabilization method akin to DAVLL, but based on a distinct physical principle: detecting the absorption signal of linearly polarized light caused by alignment, rather than circular refraction caused by orientation. This approach avoids the main drawbacks of DAVLL and its modifications.

Methods of laser frequency stabilization can be divided into those that use modulation of the laser frequency and those that do not use modulation [20]. The presence of such components prevents the use of laser light in tasks requiring high-precision spectroscopy (in particular, nonlinear sub-Doppler or Raman), as well as in a number of practical applications. In addition, a significant part of the laser power (which is usually limited in the case of semiconductor lasers) is lost in the side components of the spectrum. At the same time, methods of modulation of laser radiation frequency that use external modulators usually turn out to be disproportionately complicated, expensive, and unreliable.

To date, many schemes without modulation have been developed; all of them are based on polarization spectroscopy methods. In the 1970s it was shown [21] that the circular anisotropy signal of saturated absorption in a weak magnetic field can be used to stabilize the laser frequency. This method was applied in [22,23]; its advantage is high stability, while its disadvantages include a complex two-beam scheme, the impossibility of frequency locking outside the narrow line, and dependence on the stability of the magnetic field.

The DAVLL method [24,25] uses the circular dichroism signal in a strong magnetic field. The main advantage of the method is its simplicity, the compactness of the single-beam optical scheme, and the possibility of suppressing the laser intensity noise in a balanced way. Its main disadvantages are signal registration at DC frequency and the necessity of creating a strong magnetic field.

The DAVLL scheme can be easily modified to detect the circular birefringence signal or the sum of birefringence and dichroism signals [22,26]. The laser frequency can be tuned within the optical linewidth by rotating the balanced polarization-sensitive detector and the quarter-wave plate. Disadvantages of this scheme include the drift of the DAVLL reference frequency caused by any instability in the temperature of the reference cell [27], the polarization instability of the optical scheme, and the temperature instability of the magnetic field.

In applications where higher stability is required, the DAVLL circuit may use saturated absorption resonances [28,29]. The saturated absorption signals have a high conversion slope and small width, ensuring minimal frequency drift. However, the two-beam scheme cannot provide compactness; in addition, its range of continuous frequency tuning is extremely small.

The development of the DAVLL method using both saturated absorption resonances and saturated reflection signals is described in [28–30]. The use of millimeter-sized compact cells in [31,32] allowed development of compact devices. In [33], a three-beam scheme for detection of sub-Doppler circular dichroism resonances is presented; it is characterized by high complexity while showing high stability and low drift of the reference frequency (0.25 MHz and 0.02 MHz over a time of 10 hours, respectively).

Among the unsolved problems is the task of tuning the frequency of the reference scheme within wide limits. This task is somewhat self-contradictory, since a wide frequency tuning range means a large width of the reference frequency line, and therefore a low conversion slope, which corresponds to a low sensitivity to detuning and a high level of drift. Attempts have been made to solve this problem by using an acousto-optical modulator (AOM) in the DAVLL circuit to lock to the frequency harmonics [34], or by increasing the width of the absorption line by increasing the cell temperature [35] or reducing its size [36,37]

In [38], linearly polarized light was directed transversely to a strong magnetic field. The linear birefringence signal, which has a dispersive form suitable for stabilization, was recorded with a balance detector (t-DAVLL method). Further, in [39], the saturated absorption signal of birefringence in a magnetic field of the order of millitesla units was registered by this method. The disadvantages of this approach are the strong magnetic field and DC signal detection.

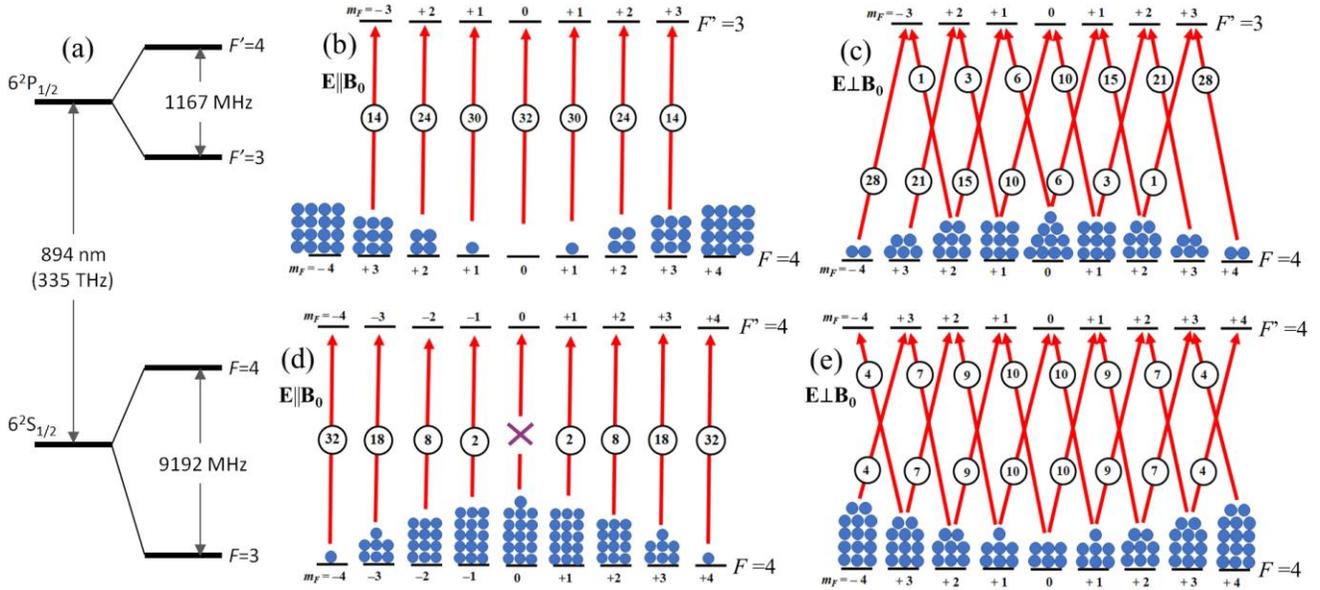

Fig. 1. (a) Scheme of the cesium $D_1$ line. (b)-(e) Schemes of the cesium line $D_1$ transitions from level $F = 4$ of the ground state in a transverse magnetic field under the influence of linearly polarized radiation: (b), (c) transition $F = 4 \to F' = 3$; (d), (e) transition $F = 4 \to F' = 4$; (b), (d) - light is polarized along the magnetic field; (c), (e) - light is polarized across the magnetic field. Relative probabilities of transitions are indicated by numbers in the circles.

To the best of our knowledge, the work closest to ours is [40], where linear dichroism signals in a cesium vacuum cell were observed in a transverse magnetic field of about 0.68 mT under $D_2$ line excitation. The polarization direction was modulated along and across the magnetic field, allowing the absorption coefficients of orthogonal polarizations to be compared. The absorption signal at the modulation frequency was recorded. The polarization modulation was carried out within very small limits, of the order of milliradian units. The use of the obtained linear dichroism signal to stabilize the laser frequency was not proposed.

Note that all methods based on the observation of dichroism in the longitudinal field allow one to obtain the error signal only at a low (conditionally "zero" or "DC") frequency, at which the laser noise is particularly high. To transfer the signal to the region of non-zero frequencies, additional devices must be used – e.g., a mechanical optical chopper. This severely limits the scope of application of such methods – as it does the use of strong magnetic fields in the original DAVLL method.

In contrast to the above works, the present work considers a method of polarization spectroscopy in a small, but non-zero transverse magnetic field, which allows stabilization of the laser frequency without its modulation. We propose our own variant of the DAVLL scheme in a modulated transverse magnetic field – *Transverse Linear Dichroic Atomic Vapor Laser Lock* (*TL-DAVLL*)[1]. As will be shown below, this method is free from the abovementioned drawbacks, while being *i*) extremely simple and *ii*) potentially providing both high resolution at the level of hundreds or even tens of kilohertz, and high resilience to variation of parameters (such as cell temperature, laser intensity, etc.).

This article is organized as follows: in Section II, we outline the basic principles of the effect of linear dichroism in the optically pumped atomic media and define the direction of the research. In Section III, we describe the experimental setup. In Section IV, we present the results of investigation of the parameters of the proposed TL-DAVLL method and estimate the achievable resolution and sensitivity; we also present the results of a laser frequency locking demo experiment. In Section V, we discuss the main advantages and disadvantages of the method.

## II. BASIC PRINCIPLES

Let us consider the interaction of an alkaline atom characterized by nuclear spin $I$ with linearly polarized light directed across a constant magnetic field $\mathbf{B_0}$. The ground state of the hydrogen-like atom is split into two hyperfine sublevels $F = I – 1/2$ and $F = I + ½$ (Fig. 1a). Each hyperfine level, in turn, is split into $2F + 1$ magnetic, or Zeeman sublevels, characterized by a magnetic number $m_F$ ($m_F = –F..F$).

---

[1] Since the method involves detecting an absorption signal, this abbreviation can also stand for *Transmission Loss DAVLL*.

Let the linearly polarized light propagates along the axis $z$ (Fig. 1b,d). When the polarization vector is directed parallel to $\mathbf{B_0}$, light causes transitions at which the magnetic quantum numbers of the ground $m_F$ and excited $m_{F'}$ states do not change ($\Delta m_F = 0$). Light polarized perpendicularly to $\mathbf{B_0}$ excites transitions with $\Delta m_F = \pm 1$ (Fig. 1c,e). For the case of excitation of the $F=4$ level of cesium by linearly polarized $D_1$ line light, the relative probabilities of optical transitions are shown in Fig. 1b-e. The populations of Zeeman sublevels, determined by the balance between the optical pumping and relaxation rates, are also schematically shown. In the case of linear light polarization, the distribution of populations is symmetric. Let us represent the population distribution as a sum of $2F + 1$ multipole moments $A_l$:

$$A_l = \sum_{m_F} (-1)^{F-m_F} \langle F, M, F, -M; l, 0 \rangle n_{m_F}, \quad (1)$$

where $\langle F, M, F, -M; l, 0 \rangle$ is the Clebsch-Gordon coefficient, $n_{mF}$ is the population of Zeeman $m_F$ level of the state $F$. Multipole moments (2) represent a particular case of the decomposition of the medium density matrix by irreducible tensor components in the so-called (κq)-representation in case of $q = 0$ [41]. Linearly polarized light produces a symmetric population distribution described by the sum of moments with even values of $l$, the smallest of which is the quadrupole ($l = 2$) moment called the alignment. The non-isotropic absorption of light is proportional to the product of the alignment and the corresponding moment of the polarization matrix of the light [41].

In cells filled with buffer gas at pressures of several torr and higher, pumping occurs by the mechanism of "depopulation" [42]: during the typical time (~30 ns) that an atom spends in the excited state, the populations of the excited state sublevels almost completely equalize due to frequent collisions with buffer gas atoms (or molecules). In this simplest case the changes in the population of the ground state levels are inversely proportional to the optical transitions probabilities (indicated by the numbers in the circles in Fig. 1) [42].

Fig. 1 shows that $i$) under identical conditions, the signs of the alignment at each of the two transitions $F = 4 \rightarrow F' = 4$ and $F = 4 \rightarrow F' = 3$ are opposite; $ii$) the alignment at each transition changes sign when the polarization orientation with respect to the magnetic field changes. The absorption coefficients of such an optically oriented (more precisely, *aligned*) medium appear to be different for the detecting light with different polarizations. When the pumping light also acts as the detecting light, it turns out that, regardless of the sign of the alignment (and hence of the polarization direction), the light bleaches the medium, and the sign of this effect is always the same – the absorption coefficient of the medium decreases with increasing pump efficiency. But if we consider the alignment signal (i.e. the difference $\Delta I$ of the intensities of the beam passed through the medium at $\mathbf{E} \| \mathbf{B_0}$ and $\mathbf{E} \perp \mathbf{B_0}$) in the system, it turns out to be sign-variable under certain conditions (as our experiment showed, one of these conditions is sufficient overlap of optical profiles corresponding to transitions $F = 4 \rightarrow F' = 4$ and $F = 4 \rightarrow F' = 3$). The alignment signal parameters are determined by many factors, but above all by the laser frequency $f$. Periodically switching the direction of the light polarization with respect to the magnetic field allows this signal to be shifted to a non-zero frequency.

A significant simplification of the optical scheme can be achieved by switching the direction of the transverse magnetic field (Fig. 2b) in the plane perpendicular to the laser beam to a direction perpendicular to the original one. If such modulation is carried out at a rate not much higher than the relaxation rate of the ground state sublevels, the intensity of the radiation passing through the cell is modulated at the field switching frequency. The transverse dichroism signal $S_{TI}$ can be extracted from the time dependence of the intensity by lock-in amplification of $\Delta I$ at the field switching frequency.

It follows from the above that both the value of $S_{TI}$ and its sign depend on the laser frequency $f$. This makes it possible to use the difference signal $S_{TI}$ to stabilize the frequency $f$, which will be demonstrated below. In particular, when pumping a low-frequency pair of optical $D_1$ transitions of the Cs line in a cell with a buffer gas, the value of $S_{TI}$ goes through zero and changes its sign while the laser frequency sweeps from the transition $F = 4 \rightarrow F' = 3$ to $F = 4 \rightarrow F' = 4$. Note that the overlapping of the optical profiles of the interrogated transitions is essential for this purpose: in a vacuum cell where very strong alignment signals can be observed, the change of the sign of $S_{TI}$ does not occur.

The physical implementation of the proposed method is as follows (Fig. 2): a cell containing an alkali metal and a buffer gas is placed inside a magnetic shield and heated to a temperature providing an optimal optical density of atomic vapor. A two- or three-dimensional system of Helmholtz coils is placed around the cell inside the shield. Stabilized laser radiation is passed through the cell (we assume that the beam direction coincides with the z-axis and the polarization direction of the beam coincides with the x-axis). A voltage is alternately applied to the $x$ and $y$ Helmholtz coils, creating a magnetic field of the order of units or tens of microtesla (the larger the field, the more stable the system is to external perturbations, and the lower the shielding requirements). The $x$, $y$, and $z$ coils can also be used to compensate for external magnetic fields. The intensity of the light transmitted through the cell is measured by a photodetector. As shown above, different directions of the transverse magnetic field correspond to different absorption coefficients (the so-called linear dichroism effect); this difference is measured by lock-in amplification of the transmitted radiation signal, with the signal controlling one of the coils used as a reference signal. The magnitude and sign of this difference depends on the detuning of the laser radiation frequency relative to the atomic transition, so a feedback loop can be used to lock the laser frequency to the frequency approximately corresponding to the maximum of the absorption profile in the cell. An additional photodetector can be used to detect laser intensity noise for subsequent subtraction from the signal.

As will be shown below, the time dependence of the intensity contains not only the first harmonic but also a strong second harmonic (the transverse dichroism signal $S_{T2}$), which can also be used as the error signal. In our experiment we used cesium as the alkaline metal, but the proposed method can be extended to other alkali metals, primarily rubidium.

### III. EXPERIMENTAL SETUP

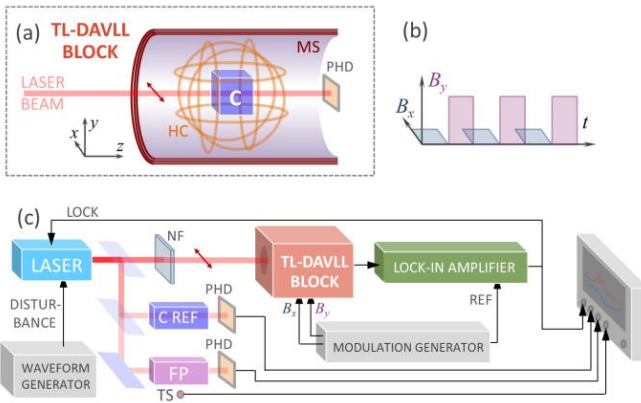

Fig. 2. Experimental setup: (a) diagram of the TL-DAVLL unit, (b) timing diagram of magnetic field switching, (c) diagram of the experiment to study the parameters of the TL-DAVLL unit. LASER – VitaWave External Cavity Diode Laser 895 nm, C – Cs+$N_2$ cell, NF – neutral filter, MS – magnetic shield, HC – Helmholtz coils, C REF – reference Cs cell, NF – neutral filter, FP – Fabry-Perot interferometer, PHD – photodiodes.

A scheme of the experimental setup is shown in Fig. 2. A cell with an internal size of $5\times5\times5$ mm$^3$ containing a few milligrams of Cs and nitrogen at a pressure of 200 torr was placed in a thermostat (not shown in the scheme) and in a cylindrical three-layer magnetic shield [43,44]. The dependences of the cell transmittance signal $\Delta I$ on the detuning $\Delta f$ of the laser radiation from the Cs absorption lines were studied. The laser power $P_p$, the cell temperature $T$, the frequency $f_{mod}$ and amplitude $B_{mod}$ of modulation, as well as the value of the longitudinal field $B_z$ were varied in the wide range. At each parameter's value, the laser frequency was scanned in the vicinity of the absorption lines $F = 4 \leftrightarrow F' = 3, F' = 4$ (low-frequency pair) or $F = 3 \leftrightarrow F' = 3, F' = 4$ (high-frequency pair), and the dependence $\Delta I (\Delta f)$ was recorded. Simultaneously, the intensity of the light that passed through the reference vacuum cell was recorded. The range of continuous frequency tuning of the laser radiation was 8–10 GHz. The signal from the reference cell was approximated by the sum of two Gaussian profiles on a nonlinear substrate; thus, the frequency scale of each recording was determined.

Linearly polarized light with a wavelength of 895 nm ($D_1$ line of Cs) generated by an external cavity semiconductor laser (VitaWave Company) was transmitted through the cell in the direction of the shield axis (z-axis). The beam cross-section was ~3 mm$^2$. A part of the laser beam was split off by a semi-transparent mirror and directed to a Cs vapor reference cell. The light passed through the cells was registered by photodiodes.

The transverse field with modulated orientation (x-y) was generated by a Helmholtz coils 3D system; the modulation frequency $f_{mod}$ was varied between 20 Hz and 20 kHz. The longitudinal (z) pair of coils was used to compensate the residual longitudinal field and to create a test field along the z axis.

The main cell transmittance signal was amplified by a low-noise amplifier and detected and filtered by two Stanford Research SR830 lock-in amplifiers at the first and second harmonics of $f_{mod}$. Next, the resulting error signal was numerically filtered; the locking frequency (at which the signal takes zero value) and the steepness of the signal at this point were calculated. All the signals were recalculated to the photocurrent level at the photodetector (Hamamatsu photodiode, efficiency of conversion of radiation power into photocurrent ~0.7 A/W).

Measurements (unless otherwise specified) were performed at $B_{mod} = 3$ μT and $f_{mod} = 130$ Hz.

### IV. RESULTS

Fig. 3 shows examples of signal dependences on the laser frequency detuning from the $F = 4 \leftrightarrow F' = 3$ transition both in the reference vacuum cell (taken at room temperature) (a), and in the TL-DAVLL cell (taken at T=95 °C) (b),(c). As can be seen from Fig. 3b, at this temperature the light absorption in the center of the line is very high at low incident light intensities. But as the pumping power increases, the medium bleaches and the dichroism signals $S_{T1}$ (Fig. 3c) at the first harmonic of the modulation frequency are detected with very good signal-to-noise ratio (signals $S_{T2}$ at the second harmonic demonstrate similar frequency dependences with an amplitude that is almost an order of magnitude smaller). As expected, the dichroism signals change their sign in the region between the transition frequencies $F=4 \leftrightarrow F'=3$ and $F = 4 \leftrightarrow F' = 4$ in the gas cell (a collisional shift of the optical transitions in the gas cell is calculated using data [45]; the positions of optical profiles displaced by collisions are shown in Fig. 3c with dotted lines).

It should be noted that the width of the linear part of the $S_{T1}$ dependence on the frequency tuning (Fig. 3c) allows, with some loss of accuracy, a controlled tuning of the laser frequency within the profile of the absorption line. This can be achieved by introducing an artificial offset to the value of the locking threshold. Let the signal $S_{T1}(v)$ steepness be $K = dS_{T1}/dv$. When stabilizing with a DC offset of magnitude $A$, the frequency shift $\Delta v$ relative to the zero point of the $S_{T1}(v)$ signal is $\Delta v = A/K$. When the slope changes by $dK$, the relative change in frequency will be $d\Delta v/\Delta v \approx dK/K$. In other words, the relative change in the frequency shift is equal to the relative change in the slope of the frequency dependence of the signal. Thus, even with $\Delta v = 100$ MHz and $dK/K = 0.1$ (a clearly overestimated value), the frequency variation will be 10 MHz, which is quite acceptable in many applications.

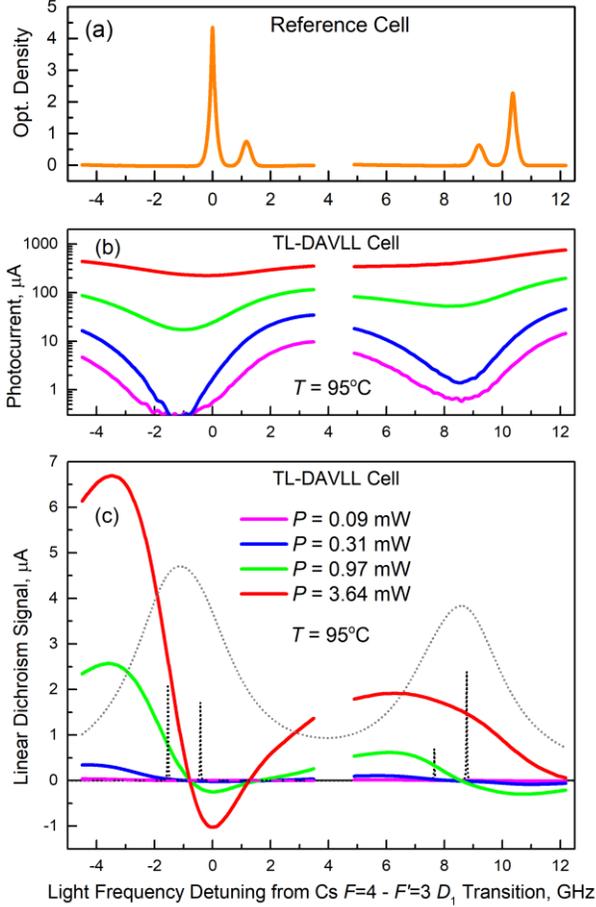

Fig. 3. Examples of signal dependences in the cells on the laser frequency detuning from the transition $F = 4 \leftrightarrow F' = 3$ in the reference vacuum cell: (a) transmittance signal converted to optical density in the reference cell; (b) transmittance signal (in photocurrent units) in the TL-DAVLL cell at different incident light intensities at $B_{mod} = 3$ μT; (c) dichroism signal at the first harmonic $S_{T1}$ (in photocurrent units) in the TL-DAVLL cell at different incident light intensities. The colors in (b) and in (c) are the same. The light gray dotted lines in (c) show the calculated absorption profiles in the gas cell taking into account the Doppler broadening and collision broadening and shift, the dark gray dotted lines – without broadening.

Parametric dependences of the frequency at which the laser is stabilized (in the absence of an artificial offset) using the first and second harmonics are shown in Fig. 4a, b. The corresponding values of the steepness of the dichroism signals in the vicinity of the transition $F = 4 \leftrightarrow F' = 3$ are shown in Fig. 4c,d. It can be seen that the above dependences have extrema both in temperature and pumping power.

Now we evaluate the ultimate parameters of the method assuming that the technical noise of the laser at the modulation frequency is suppressed to the level of the shot noise of the photocurrent (this is easier to achieve if the modulation frequency is high enough):

$$\delta f = \sqrt{2e \cdot I_{PH}} \cdot \left( \left. \frac{dS_d}{df} \right|_{S_D=0} \right)^{-1}. \quad (2)$$

In Fig. 5a,b the dependences given in Fig. 4a,c are presented in the form of contour maps. The maximum value of the achievable resolution in Fig. 6c is ~2 kHz/√Hz, but it is realized under sub-optimal conditions in terms of stability. Fig. 5c shows the estimated value of the maximum achievable resolution limited by the shot noise. It was calculated by dividing the shot noise value calculated from the experimentally measured photocurrent by the experimentally measured steepness value.

The white cross in Fig. 5a-c marks the point where the minima of the temperature and pumping power dependence of the reference frequency are reached and the corresponding first derivatives are zeroed. In this case, the quadratic coefficients of the dependence of the stabilization frequency on the cell temperature and the power of the incoming radiation measured at $T$=90°C, $P_p$=1.0 mW were as follows

$$d^2f/dI^2 = -180 \pm 30 \text{ MHz/mW}^2,$$
$$d^2f/dT^2 = -0.61 \pm 0.29 \text{ MHz/°C}^2. \quad (3)$$

All similar dependences were measured for the dichroism signals in the vicinity of transitions $F = 3 \leftrightarrow F' = 3, F' = 4$. It turned out that the steepness of these signals is about an order of magnitude less than the steepness of the signals near the $F = 4 \leftrightarrow F' = 3, F' = 4$ transitions. Furthermore, the zero crossing of the dichroism signal is not reached in all pumping regimes. Nevertheless, local extrema of the frequency dependence on temperature and pumping power are also observed for the high-frequency pair of transitions.

The dependence of the frequency on the steepness of the dichroism signals was somewhat unexpected (Fig. 6). One could expect that the frequency decay of the steepness starts at $f_{mod} > \Gamma/2\pi$, where $\Gamma/2\pi$ is the relaxation rate of the Cs ground state sublevels in the cell under study (in our experiment $\Gamma/2\pi$ varied from 350 Hz to 2 kHz depending on the pumping parameters), but the dependences we obtained turned out to be much more complicated.

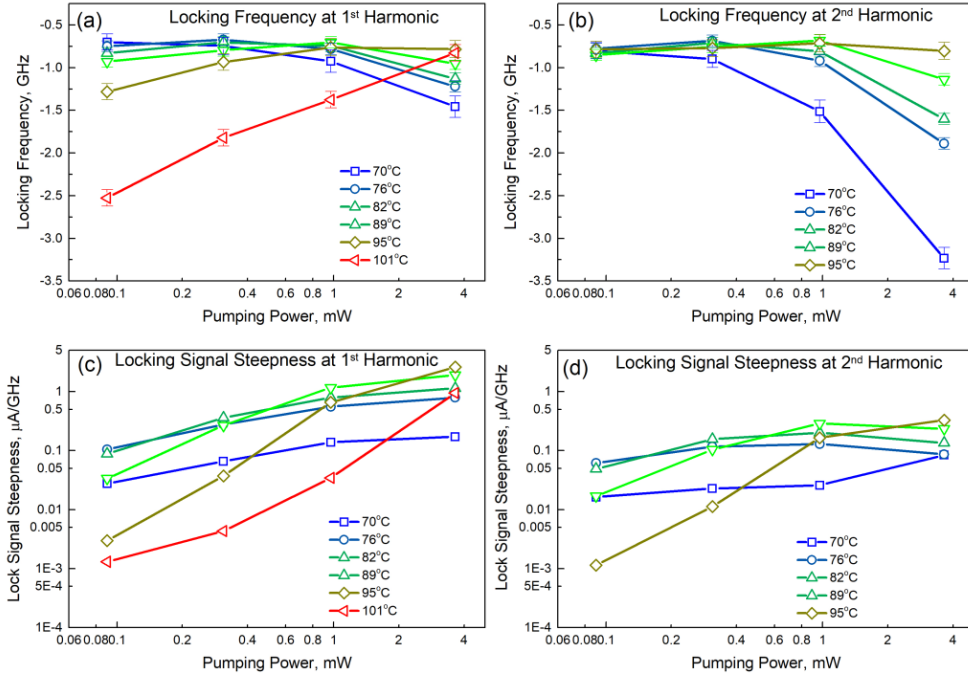

Fig. 4. Dependences of the locking frequency (a), (b) and locking signal steepness (c), (d) of the dichroism signals in the vicinity of the transition $F = 4 \leftrightarrow F' = 3$ at the first (a), (c) and second (b), (d) harmonics at different temperatures as a function of pumping power. The dependences were taken at $f_{mod} = 130$ Hz.

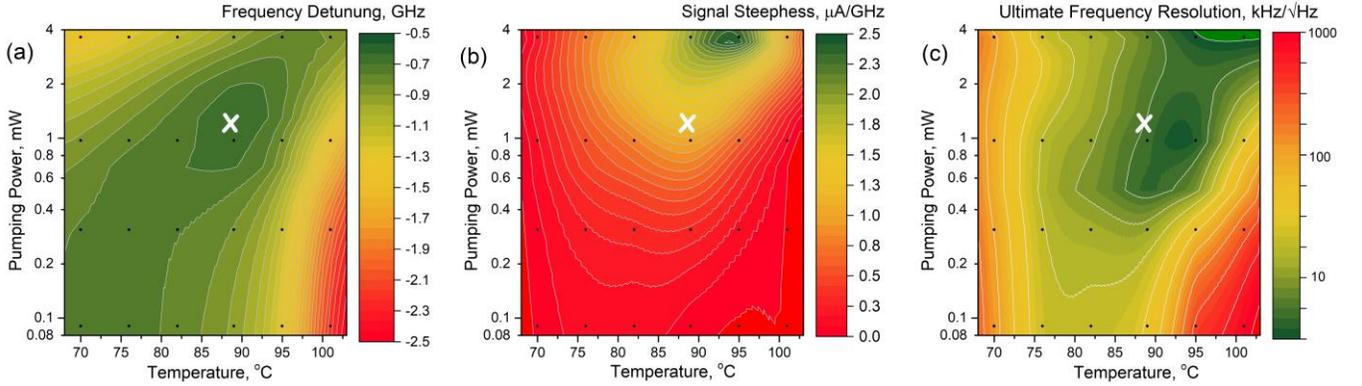

Fig. 5. Dependences of (a) the locking frequency in the vicinity of the transition $F = 4 \leftrightarrow F' = 3$, (b) the corresponding steepness of the dichroism signals, and (c) the maximum achievable resolution limited by shot noise on the cell temperature and pumping power. The dependences were recorded at $f_{mod} = 130$ Hz.

Obviously, at high ($> \Gamma/2\pi$) frequencies the dynamics of the system when switching the direction of the magnetic field is determined not by relaxation, but by re-distribution of the second-order magnetic moments to new magnetic field directions. In our case, the system proved to be operable at frequencies up to $f_{mod} \approx 10$ kHz. At the frequency $f_{mod} \approx 3.5$ kHz the phase inversion of the signals occurs; it is noteworthy that these frequencies are almost independent of the magnitude of the field generated by the transverse coils. All these effects require further investigation, but it is beyond the scope of this article, devoted to the description of the method and demonstration of its performance.

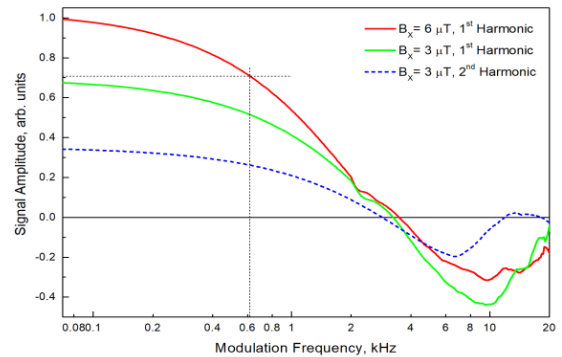

Fig. 6. Frequency dependences of the dichroism signal steepness in the vicinity of the $F = 4 \leftrightarrow F' = 3$ transition.

Another parameter that can affect the performance of the circuit is the presence of the residual longitudinal magnetic field. We studied its influence and found that as long as the longitudinal field does not exceed ~20% of the transverse field (0.8 μT in our case), it has no noticeable effect on the reference frequency (Fig. 7a). The steepness of the signal decreases by a factor of less than $1/\sqrt{2}$ in this case (Fig. 7b). This result is very significant for practical applications: it means that the proposed method not only does not use strong magnetic fields, but also does not require particularly good magnetic shielding.

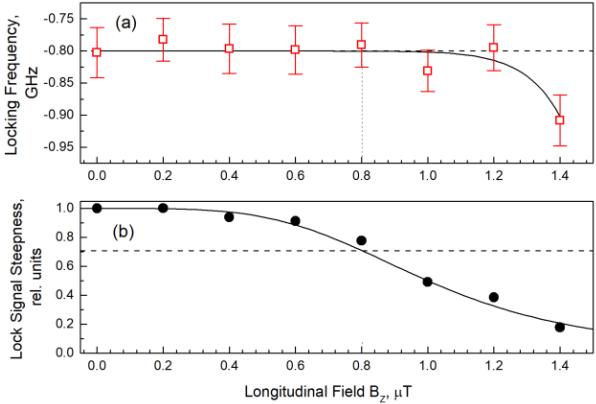

Fig. 7. Dependences of (a) the reference frequency and (b) the steepness of the dichroism signals on the longitudinal magnetic field in the vicinity of the transition $F = 4 \leftrightarrow F' = 3$ at $T = 90$ °C, $I=1.0$ mW, and $f_{mod} = 130$ Hz.

We tried to perform a demo stabilization of the laser using the transverse dichroism signal. To monitor the laser frequency we used a Fabry-Perot SA-200 interferometer (Thorlabs). It should be noted that temperature variations of the interferometer frequency dominate in the scheme – the temperature coefficient of invar expansion ($1.2 \cdot 10^{-6}$ K$^{-1}$) corresponds to the temperature coefficient of the interferometer frequency ~0.3 GHz/K, or 30 MHz at temperature change $dT = 0.1$ °C.
Therefore, we first artificially expanded the interferometer bandwidth (~7.5 MHz) by switching the interferometer photodiode into a significantly nonlinear photovoltaic mode – in this case, the width of the quasi-linear part of the interferometer output signal characteristic was 75 MHz. Second, we had to add independent temperature control of the interferometer body to the circuit. Third, we had to demonstrate the performance of the laser frequency stabilization scheme by introducing artificial perturbations into the laser frequency, similar to what was done in [37]. In our case, a sinusoidal perturbation at a frequency of 0.1 Hz was introduced into the laser diode current, and the detected error signal from the TL-DAVLL block was fed to the piezoceramic control input of the laser external resonator.

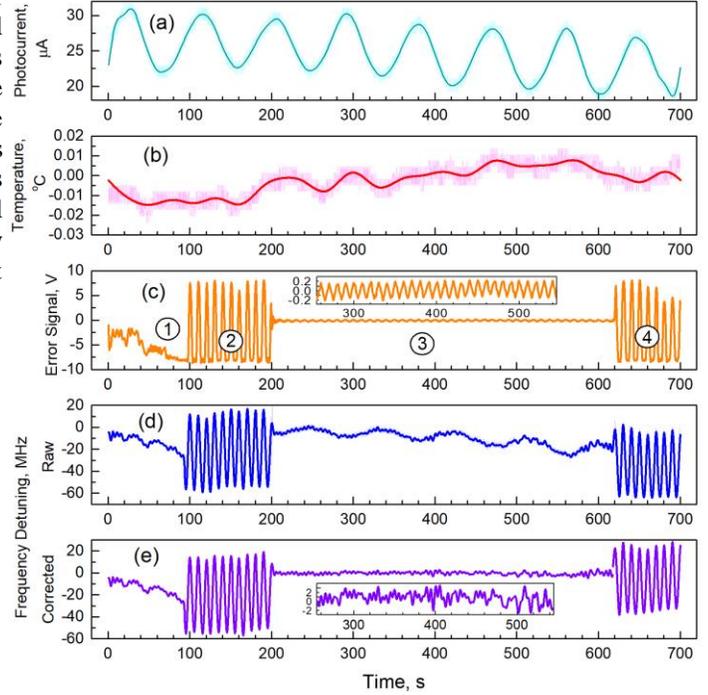

Fig. 8. Results of an experiment demonstrating the suppression of external laser frequency perturbations using a stabilization scheme: (a) photocurrent intensity at the output of the TL-DAVLL unit; (b) air temperature variation in the vicinity of the Fabry-Perot interferometer; (c) signal from the output of the SR830 lock-in amplifier (error signal); (d) signal from the output of the Fabry-Perot interferometer converted to the frequency scale; (e) the same signal after subtracting the quadratic long-term fit drift and correlations associated with the photocurrent intensity. The insets show enlarged fragments of the corresponding recordings. Time intervals: (1) – laser frequency stabilization is off, laser modulation is off; (2), (4) – stabilization is off, modulation is on; (3) – stabilization is on, modulation is on.

Fig. 8 shows the results of a demonstration experiment performed at $f_{mod} = 1$ kHz. No balanced photodetector was used in the experiment, to demonstrate the performance of the simplest version of the TL-DAVLL block. Therefore, the modulation frequency was increased to reduce the influence of low-frequency laser noise.

As can be seen from Fig. 8a, the temperature controller of the TL-DAVLL cell operated in the oscillating mode; the magnitude of slow oscillations was ~1.2 °C, which, given the circuit parameters, resulted in frequency variations of ~7.7 MHz (Fig. 8d). The interferometer temperature drifted slowly within 0.02 °C (Fig. 8b), corresponding to a frequency drift of about 6 MHz (Fig. 8d).

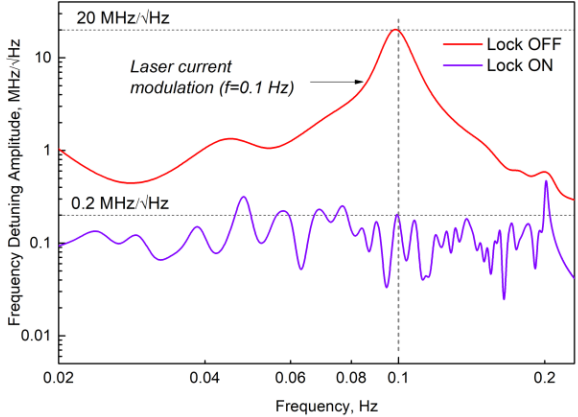

Fig. 9. Interferometer signal spectra with the stabilization loop turned off and on (see Fig. 8).

At the interval 0–200 s the stabilization scheme was switched off. At t = 100 s, slow modulation of the laser current with a frequency of 0.1 Hz was turned on. At t = 200 s, the stabilization scheme was turned on and it was turned off at t = 610 s. At this interval, the 0.1 Hz component disappears from the interferometer signal but appears in the error signal of the TL-DAVLL circuit (Fig. 8c).
During processing, we subtracted long-term drifts, specifically, the component correlated with the cell temperature (Fig. 8a) and the quadratic baseline representing the total drift from the interferometer signal (Fig. 8b). The result is shown Fig. 8e. Note that subtracting slow components did not affect the calculation of variations at 0.1 Hz. Figure 9 demonstrates the suppression of external perturbations by comparing the spectra of the interferometer signal with the stabilization loop turned off and on.

It can be seen that the perturbation is suppressed by at least two orders of magnitude. The time constant of the Stanford Research SR830 lock-in amplifier was 10 s at a gain drop of 6 dB/oct. Taking into consideration a suppression factor in the feedback loop of K ≈ 100 (Fig. 8c, Fig. 9), it corresponds to a time constant $\tau \approx 0.1$ s at closed-loop feedback. Further increase of the control coefficient is possible by using a PID controller.

## V. DISCUSSION

We investigated the parameters of our proposed TL-DAVLL method, evaluated its ultimate parameters, and performed an experiment demonstrating efficiency of the method. The suppression of artificially introduced interference up to 0.1–0.2 MHz/√Hz was achieved in a demonstration experiment (Fig. 9).

Based on the ultimate parameters of the scheme obtained above (3), we can estimate the achievable level of its drift. Let us assume that the uncontrolled variations of the parameters in the system are $dI/I = 0.01$ and $dT = 0.1$ °C. Substituting these values into (3), we obtain a total frequency instability of 24 kHz. The ultimate (shot noise limited) frequency resolution under the same conditions is 3.9 kHz/√Hz. Thus, at the full width of the optical absorption profile in the cell of ~3.2 GHz, the relative ultimate accuracy of the method reaches $8 \cdot 10^{-6}$.

At the same time, with a constant temperature setting error of only $\Delta T = 1$ °C and the same level of variation $dT = 0.1$ °C, the total frequency instability reaches 0.1 MHz, and this value grows quadratically with $\Delta T$.

In addition, it should be taken into account that, due to the complicated dynamics of the alignment momentum, the scheme parameters change with the modulation frequency; thus, in our demonstration experiment, changing $f_{mod}$ from 130 Hz to 1 kHz led to a shift of the extremum of the temperature dependence, and to the appearance of a linear temperature coefficient of ~6.5 MHz/°C (Fig. 8(c)). However, the modulation frequency is a very well controlled scheme parameter and its variations can be ignored.

The obtained results allow us to draw a number of conclusions regarding both the proposed method in general and its specific application for stabilization of laser radiation by the $D_1$ absorption line of Cs.

General conclusions:
i) The method is extremely simple: the setup for the error signal detection consists of a magnetic shield, a 2D or 3D Helmholtz coil, a cell, a heater and a photodetector;
ii) the method does not require any type of modulation of the laser beam;
iii) the modulation of the magnetic field allows the dichroism signal at a non-zero frequency to be registered;
iv) the method does not require strong magnetic fields; it is realized in transverse fields in µT range;
v) the resilience of the parameters to residual longitudinal fields at the level of tenths of µT allows to use minimal shielding, like single-layer shield or even a system of compensating coils;
vi) the registration of the dichroism signal is possible on the second harmonic of the modulation frequency, which even in the most compact version of the scheme allows for elimination of direct interference from magnetic fields on the photodetector, cables and elements of electrical circuits.

The laser stabilization frequency in this method is determined by the collisional shift of transition frequencies in the used cell; therefore, the choice of the locking frequency can be made not only by introducing the locking level artificial offset, but also by selecting the gas filling of the cell.

Below we present conclusions with respect to the concrete application of the method for stabilization of laser radiation by the Cs $D_1$ absorption line in the vicinity of the $F = 4 \leftrightarrow F' = 3$ transition:
i) stabilization of the laser using the zero point of the dichroism signal is possible at the frequency

corresponding approximately to the top of the absorption profile in the gas cell;

*ii)* high achievable signal-to-noise ratio ($\sim 10^6$ with respect to the shot noise) makes it possible to realize a frequency resolution at the level of kilohertz or tens of kilohertz in one hertz bandwidth;

*iii)* the presence of a smooth extremum of the dependence of the locking frequency on the cell temperature and light intensity makes it possible to achieve high (at the level of hundreds or even tens of kilohertz) stability of the locking frequency;

*iv)* the width of the linear part of the dependence of the dichroism signal on the frequency tuning allows (with a corresponding loss of accuracy) a controlled tuning of the laser light frequency within the profile of the absorption line (depending on the cell gas filling, from hundreds of MHz to units of GHz);

*v)* the magnetic field modulation frequency can reach ten kilohertz.

The measured dependences (Fig. 3(c)) suggest that the shape of the dichroism signal dependence on frequency tuning also allows stabilization of the laser outside the absorption line at a distance of up to several line widths from the lowest-frequency transition of the $D_1$ line, which can be used for quantum nondestructive measurement.

## VI. CONCLUSIONS

We have presented a method, similar to the DAVLL method – Transverse Linear Dichroic Atomic Vapor Laser Lock (TL-DAVLL), which, as far as can be judged from the results of our studies, allows us to avoid the main disadvantages of the existing schemes. The method is based on the dichroism signals in the transverse magnetic field, which were studied with a laser tuned to the $D_1$ line of Cs in a gas-filled cell. We have demonstrated applicability of this scheme for stabilization of the laser frequency both in the vicinity of the pair of low-frequency transitions. In this case, parameters of the signals were obtained which illustrate the high achievable stability and resolving power of the method. Based on the similarity of the optical pumping processes in Cs and $^{87}$Rb, we believe that there is every reason to expect that the method will also be applicable to the $D_1$ line of $^{87}$Rb.

## VII. CONFLICT OF INTEREST.

The authors declare that they have no conflict of interest.

## VIII. ACKNOWLEDGEMENTS

The authors thank Valerii S. Zapasskii for attracting interest in the effects of transverse linear dichroism in alkali atoms. The authors are very grateful to Sofia K. Vershovski-Heisler for the invaluable assistance in editing the style and grammar of the text.